\documentclass[aps,twocolumn, noshowpacs, floatfix]{revtex4}
\usepackage{}

\usepackage{amsfonts}
\usepackage{amssymb}
\usepackage{graphicx}
\usepackage{amsmath}
\usepackage[english]{babel}
\usepackage{color}

\begin{document}

\title{Polaron dynamics with off-diagonal coupling: beyond the Ehrenfest approximation}

\author{Zhongkai Huang$^{1}$, Lu Wang$^{1,2}$, Changqin Wu$^{3,4}$, Lipeng Chen$^{1}$, Frank Grossmann$^{5}$, and Yang Zhao$^{1}$\footnote{Electronic address:~\url{YZhao@ntu.edu.sg}}}
\affiliation{$^1$Division of Materials Science, Nanyang Technological University, Singapore 639798, Singapore\\
$^2$Department of Physics, Zhejiang University, Hangzhou 310027, China\\
$^3$State Key Laboratory of Surface Physics and Department of Physics, Fudan University, Shanghai 200433, China\\
$^4$Collaborative Innovation Center of Advanced Microstructures, Fudan University, Shanghai 200433, China\\
$^5$Institute for Theoretical Physics, Technische Universit\"at Dresden, D-01062 Dresden, Germany}

\begin{abstract}
Treated traditionally by the Ehrenfest approximation, dynamics of a one-dimensional molecular crystal model with off-diagonal exciton-phonon coupling is investigated in this work using the Dirac-Frenkel time-dependent variational principle with the multi-D$_2$ {\it Ansatz}. It is shown that the Ehrenfest method is equivalent to our variational method with the single D$_2$ {\it Ansatz}, and with the multi-D$_2$ {\it Ansatz}, the accuracy of our simulated dynamics is significantly enhanced in comparison with the semi-classical Ehrenfest dynamics. The multi-D$_2$ {\it Ansatz} is able to capture numerically accurate exciton momentum probability and help clarify the relation between the exciton momentum redistribution and the exciton energy relaxation. The results demonstrate that the exciton momentum distributions in the steady state are determined by a combination of the transfer integral and the off-diagonal coupling strength, independent of the excitonic initial conditions. We also probe the effect of the transfer integral and the off-diagonal coupling on exciton transport in both real and reciprocal space representations. Finally, the variational method with importance sampling is employed to investigate temperature effects on exciton transport using the multi-$\rm D_2$ {\it Ansatz}, and it is demonstrated that the variational approach is valid in both low and high temperature regimes.
\end{abstract}
\date{\today}

\maketitle

\section{Introduction}
Conducting polymers (CPs) are a special class of organic materials with electronic and ionic conductivity, advanced processability and extraordinary wettability \cite{xu_2005, das_2012}. In $1977$, Heeger {\it et al.}~reported oxidized iodine-doped polyacetylene as a forerunner of CPs \cite{shir_1977, heller_2015}. Various experimental strategies have been developed to produce CPs by techniques such as monomer oxidation using chemical oxidative polymerization in solution \cite{darm_2014}, electrochemical polymerization on conductive substrates \cite{darmanin_2011}, and vapor-phase polymerization \cite{im_2008}. Electrical properties of CPs can be tuned by oxidation and reduction, giving rise to rapid growth of applications. Based on their good charge transport property and high quantum efficiency of the luminescence, important utilizations of CPs are found in the large scale organic light-emitting diodes \cite{burr_1990} and electronic devices such as field-effect transistors \cite{torsi_2000}. CPs have also been used as an electrode material for supercapacitors \cite{fusa_2001, reddy_2014}. As a logical alternative to conventional inorganic electrode materials, a composite architecture of various CPs have been developed as a cathode for ultrafast rechargeable batteries \cite{kim_2014}. In comparison with non-conducting polymers, there are many advantages of CPs with regard to their electronic properties. CPs have also been used for other purposes \cite{ates_2012}. For example, because of easy processibility in microsturing processes \cite{schultze_2005}, CPs have been considered for a wide range of biomedical and bioengineering applications: artificial muscles \cite{otero_1998}, controlled drug release \cite{abidian_2006}, and neural recording \cite{abidian_2009, ravi_2010}. Surface wettability based on CPs can switch between superhydrophilicity and superoleophobicity by surface morphology control at nanoscale \cite{darm_2014}, implying usage of CPs in intelligent orthopedic and dental implants \cite{liao_2013}.

In the aforementioned applications, the efficiency of charge carrier transport and exciton transport significantly impacts the overall device performance \cite{ates_2012}. The carrier or exciton transport in CPs is well described by the Su-Schrieffer-Heeger (SSH) model in which the $\pi$ electrons are treated in a tight-binding approximation and the $\sigma$ electrons are assumed to move adiabatically with the nuclei \cite{heeger_1988}. Su {\it et al.}~convincingly demonstrated that solitons play a critical role in the carrier transport doping mechanism \cite{su_1979}. Troisi {\it et al.}~applied the SSH model to investigate charge carrier dynamics in crystalline organic semiconductors by solving the time-dependent Schr\"odinger equation for the charge wave function and using the Ehrenfest theorem for classical accelerations of nuclear positions \cite{troisi_2006, ehrenfest_1927}. Improvements on this semi-classical method  has been made to study charge transport in organic materials in recent years \cite{hultell_2006, troisi_2009, si_2015}. Temperature dependent charge carrier mobility has also been considered \cite{troisi_2006, si_2015}. It is believed that for short times (comparable to the phonon period) the evolution of the system is dominated by semi-classical dynamics. The traditional Ehrenfest dynamics did not well treat the decoherence effect, which is incorporated by an instantaneous decoherence correction (IDC) approach in the framework of semi-classical method \cite{si_2015, dong_2016}.

Even though the semi-classical dynamics in the SSH model can capture certain features of charge transport, enormous challenges still remain to accurately describe fully quantum dynamical correlations between the electronic and vibrational subsystems \cite{guanqi_2013}. In realistic polymer chains, charge transport processes occur on the nano scale and the carriers interact with the environment including the dominant phonon degrees of freedom \cite{szymanski_2005}. The SSH model includes off-diagonal exciton-phonon coupling as a nontrivial dependence of the exciton transfer integral on lattice coordinates \cite{su_1979, zh_97}. Due to inherent difficulties, the off-diagonal coupling is often inadequately treated in theoretical studies. Early treatments of the off-diagonal coupling include the Munn-Silbey theory \cite{mu_85}. Recently, the Davydov D$_2$ {\it Ansatz} \cite{zhao_12} and the multiple Davydov trial states \cite{zhou_15} have been developed to study polaron dynamics in the presence of the off-diagonal coupling. However, much awaits to be studied on the rich polaron dynamics with off-diagonal coupling with regard to exciton momentum redistribution and energy relaxations \cite{dorfner_2015}.

In this work, In order to offer an accurate description of polaron dynamics including off-diagonal coupling, the Dirac-Frenkel time-dependent variational approach with the multiple Davydov trial states will be employed. We also aim to examine the accuracy of the Ehrenfest dynamics in the SSH model. We first demonstrate that the semi-classical method and the variational method using the single D$_2$ {\it Ansatz} are equivalent. Then we check the validity of the semi-classical method (the variational method with the single D$_2$ {\it Ansatz}) by examining its deviations from the exact quantum dynamics. The underlying physics is revealed in the real and reciprocal space representations, including the exciton transport, the exciton momentum redistribution and the exciton energy dissipation. At the closing of the paper, we show that the fully quantum mechanical method using our multiple Davydov trial states is also applicable at the finite temperatures.

The reminder of the paper is structured as follows. In Sec.~\ref{methodology}, we present the model Hamiltonian and the variational wave function, the multi-$\rm D_2$ {\it Ansatz} used for describing the exciton transport. In Sec.~\ref{Considering a one-dimensional stack of planar conjugated molecules}, the accuracy of the variational method using the multi-$\rm D_2$ {\it Ansatz} is examined by the ansatz deviation, which quantifies how faithfully
the trial state follows the Schr\"odinger equation, and it is shown that large enhancement over that of the semi-classical method have been achieved. Numerical results of polaron dynamics by the variational method using the multi-$\rm D_2$ {\it Ansatz} are discussed in Sec.~\ref{Different parameter regimes}. Impacts of the transfer integral and the off-diagonal coupling on the exciton transport are studied in Sec.~\ref{effect of transfer integral and off-diagonal coupling}. Effects of temperature on polaron dynamics is investigated in Sec.~\ref{Effect of temperature}. Conclusions are drawn in Sec.~\ref{Conclusions}.

\section{Methodology}
\label{methodology}
\subsection{Model}
\label{Model}
In presence of only off-diagonal coupling, the Hamiltonian of the one-dimensional Holstein molecular crystal model takes the form
\begin{equation}
\hat{H}=\hat{H}_{\rm ex}+\hat{H}_{\rm ph}+\hat{H}_{\rm ex-ph}^{\rm o.d.},
\label{Holstein}
\end{equation}
where $\hat{H}_{\rm ex},\hat{H}_{\rm ph}$ and $\hat{H}_{\rm ex-ph}^{\rm o.d.}$ denote the exciton Hamiltonian, the bath (phonon) Hamiltonian, and the off-diagonal exciton-phonon coupling Hamiltonian, respectively. In the site representation,
\begin{eqnarray}\label{Hamiltonian_site}
\hat{H}_{\rm ex}&=&-J\sum_{n}a_{n}^{\dagger}\left(a_{n+1}+a_{n-1}\right),\nonumber \\
\hat{H}_{\rm ph}&=&\omega_{0}\sum_{n}b_{n}^{\dagger}b_{n},\nonumber \\
\hat{H}_{\rm ex-ph}^{\rm o.d}	&=&	\frac{1}{2}\phi\omega_{0}\sum_{n,l}\left[a_{n}^{\dagger}a_{n+1}\left(b_{l}+b_{l}^{\dagger}\right)\left(\delta_{n+1,l}-\delta_{n,l}\right)\right.\nonumber \\
&  &+\left.a_{n}^{\dagger}a_{n-1}\left(b_{l}+b_{l}^{\dagger}\right)\left(\delta_{n,l}-\delta_{n-1,l}\right)\right],
\label{anti}
\end{eqnarray}
where $\hat{a}_n^{\dag}$ ($\hat{a}_n$) and $\hat{b}_n^{\dag}$ ($\hat{b}_n$) are the exciton and phonon creation (annihilation) operators for the $n$-th site, respectively. In this work, only the anti-symmetric exciton-phonon coupling is considered in Eq.~(\ref{anti}). In the phonon momentum space, we can rewrite $\hat{H}_{\rm ph}$ and $\hat{H}_{\rm ex-ph}^{\rm o.d.}$ as,
\begin{eqnarray}\label{Hamiltonian}
\hat{H}_{\rm ph} & = & \sum_{q}\omega_{q}\hat{b}_{q}^{\dagger}\hat{b}_{q},  \nonumber \\
\hat{H}_{\rm ex-ph}^{\rm o.d.} & = & \frac{1}{2}N^{-1/2}\phi\sum_{n,q}\omega_{q}\{\hat{a}_{n}^{\dagger}\hat{a}_{n+1}[e^{iqn}(e^{iq}-1)\hat{b}_{q}+ {\rm H.c.}] \nonumber \\
&  & +\hat{a}_{n}^{\dagger}\hat{a}_{n-1}[e^{iqn}(1-e^{-iq})\hat{b}_{q}+ {\rm H.c.}]\},
\end{eqnarray}
where $\omega_q$ is the phonon frequency at the phonon momentum $q$, and $\hat{b}_q^{\dag}$ ($\hat{b}_q$) is the creation (annihilation) operator of a phonon with the momentum $q$,
\begin{equation}
\hat{b}_q^{\dag} = N^{-1/2}\sum_n e^{iqn}\hat{b}_n^{\dag}, \quad \hat{b}_n^{\dag} =  N^{-1/2}\sum_q e^{-iqn}\hat{b}_q^{\dag}.
\label{momentum}
\end{equation}
The parameters $J$ and $\phi$ represent the transfer integral and the off-diagonal coupling strength, respectively.
A linear phonon dispersion is assumed,
\begin{equation}
\omega_{q}=\omega_{0}\left[1+(\frac{2\left|q\right|}{\pi}-1)W\right],
\end{equation}
where $\omega_0$ denotes a central phonon frequency, $W$ is a constant between $0$ and $1$, the bandwidth of the
phonon frequency is $2W\omega_0$, and
$q=2\pi l/N$ represents the momentum index with $l=-\frac{N}{2}+1, \ldots, \frac{N}{2}$.
In the rest of the paper, $\omega_0$ is set to unity as the energy unit, and a dispersionless optical phonon band with $W=0$ is used.
\subsection{Multiple Davydov trial states}

\label{Multiple Davydov trial state}
In this work, we employ the Dirac-Frenkel variational principle to obtain quantum dynamics. We use the multiple Davydov trial states with multiplicity $M$, which are essentially $M$ copies of the corresponding single Davydov Ansatz. The multi-${\rm D}_2$ {\it Ansatz} has less variational variables than the multi-${\rm D}_1$ {\it Ansatz} when a same $M$ is used, but performs better in illuminating the polaron dynamics with the off-diagonal coupling \cite{zhou_15}. The multi-${\rm D}_2$ state with the multiplicity $M$, can be written as
\begin{eqnarray}\label{D2_state}
&& \left|{\rm D_2^M}\left(t\right)\right\rangle  =  \sum_{i}^{M}\sum_{n}^N\psi_{in}\left|n\right\rangle \left|\lambda_{i}\right\rangle, \\ \nonumber
&& =\sum_{i}^{M}\sum_{n}^N\psi_{in} \hat{a}_{n}^{\dagger}\left|0\right\rangle _{\rm ex} \exp\left\{ \sum_{q}\left[\lambda_{iq}\hat{b}_{q}^{\dagger}-\lambda_{iq}^{\ast}\hat{b}_{q}\right]\right\} \left|0\right\rangle _{\rm ph},
\end{eqnarray}
where $\psi_{in}$ and $\lambda_{iq}$ are the exciton amplitudes and the phonon displacements, respectively, $n$ is the site index of the
molecular ring, and $i$ labels $i$-th $\rm D_2$ state in the coherent superposition. If $M=1$, the multi-${\rm D}_2$ {\it Ansatz} reduces to the original single Davydov $\rm D_2$ trial state. Equations of motion for the variational
parameters $\psi_{in}$ and $\lambda_{iq}$ are then derived by adopting the Dirac-Frenkel variational principle,
\begin{eqnarray}\label{eq:eom1}
\frac{d}{dt}\left(\frac{\partial L}{\partial\dot{\psi_{in}^{\ast}}}\right)-\frac{\partial L}{\partial\psi_{in}^{\ast}} & = & 0, \nonumber \\
\frac{d}{dt}\left(\frac{\partial L}{\partial\dot{\lambda_{iq}^{\ast}}}\right)-\frac{\partial L}{\partial\lambda_{iq}^{\ast}} & = & 0.
\end{eqnarray}
For the multi-$\rm D_2$ {\it Ansatz}, the Lagrangian $L$ is given as
\begin{eqnarray}
L & = & \langle {\rm D}^M_2(t)|\frac{i}{2}\frac{\overleftrightarrow{\partial}}{\partial t}- \hat{H}|{\rm D}^M_2(t)\rangle \nonumber \\
& = & \frac{i}{2}\left[ \langle {\rm D}^M_2(t)|\frac{\overrightarrow{\partial}}{\partial t}|{\rm D}^M_2(t)\rangle - \langle {\rm D}^M_2(t)|\frac{\overleftarrow{\partial}}{\partial t}|{\rm D}^M_2(t)\rangle \right] \nonumber \\
&-& \langle {\rm D}^M_2(t)|\hat{H}|{\rm D}^M_2(t)\rangle,
\label{Lagrangian_2}
\end{eqnarray}
where the first term yields
\begin{eqnarray}
 &  & \frac{i}{2}\left[ \langle {\rm D}^M_2(t)|\frac{\overrightarrow{\partial}}{\partial t}|{\rm D}^M_2(t)\rangle - \langle {\rm D}^M_2(t)|\frac{\overleftarrow{\partial}}{\partial t}|{\rm D}^M_2(t)\rangle \right] \nonumber \\
 &  &= \frac{i}{2}\sum_{i,j}^{M}\sum_{n}\left(\psi_{jn}^{\ast}\dot{\psi}_{in}-\dot{\psi}_{jn}^{\ast}\psi_{in}\right)S_{ji}\nonumber \\
 &  & \sum_{i,j}^{M}\sum_{n}\psi_{jn}^{\ast}\psi_{in}S_{ji}\sum_{q}\left[\frac{\dot{\lambda}_{jq}^{\ast}\lambda_{jq}+\lambda_{jq}^{\ast}\dot{\lambda}_{jq}}{2}\right.\nonumber \\
 &  & \left.-\frac{\dot{\lambda}_{iq}\lambda_{iq}^{\ast}+\lambda_{iq}\dot{\lambda}_{iq}^{\ast}}{2}+\lambda_{jq}^{\ast}\dot{\lambda}_{iq}-\lambda_{iq}\dot{\lambda}_{jq}^{\ast}\right],
\label{energies}
\end{eqnarray}
and the second term is
\begin{eqnarray}
 &  & \left\langle {\rm D}^M_{2}\left(t\right)\right|\hat{H}\left|{\rm D}^M_{2}\left(t\right)\right\rangle \nonumber \\
 &  &= \left\langle {\rm D}^M_{2}\left(t\right)\right|\hat{H}_{\rm ex}\left|{\rm D}^M_{2}\left(t\right)\right\rangle +\left\langle {\rm D}^M_{2}\left(t\right)\right|\hat{H}_{\rm ph}\left|{\rm D}^M_{2}\left(t\right)\right\rangle \nonumber \\
 &  &+\left\langle {\rm D}^M_{2}\left(t\right)\right|\hat{H}_{\rm ex-ph}^{\rm o.d.}\left|{\rm D}^M_{2}\left(t\right)\right\rangle. \nonumber \\
 \end{eqnarray}
Detailed derivations of the equations of motion for the variational parameters are given in Appendix \ref{Equations}, together with discussions on initial conditions and numerical details.

To quantify the accuracy of the variational dynamics based on the multiple Davydov trial states, we introduce a deviation vector $\vec{\delta}(t)$ defined as
\begin{eqnarray}
\vec{\delta}(t) & =  & \vec{\chi}(t) -\vec{\gamma}(t)  \nonumber \\
& = & \frac{\partial}{\partial t}|\Psi(t)\rangle - \frac{\partial}{\partial t}|{\rm D}^M_{2}(t)\rangle,
\label{deviation_1}
\end{eqnarray}
where the vectors $\vec{\chi}(t)$ and $\vec{\gamma}(t)$ obey the Schr\"{o}dinger equation  $\vec{\chi}(t)=\partial |\Psi(t)\rangle / \partial t = {-i}\hat{H}|\Psi(t)\rangle$ and the Dirac-Frenkel variational dynamics $\vec{\gamma}(t)=\partial |{\rm D}^M_{2}\rangle / \partial t$ in Eq.~(\ref{eq:eom1}), respectively. The deviation vector $\vec{\delta}(t)$ can be calculated as
\begin{equation}
\vec{\delta}(t) = -i\hat{H}|{\rm D}^M_{2}(t)\rangle - \frac{\partial}{\partial t}|{\rm D}^M_{2}(t)\rangle.
\label{deviation_2}
\end{equation}
Thus, the accuracy of the trial state is indicated by the amplitude of the deviation vector $\Delta(t)=||\vec{\delta}(t)||$.
In order to view the deviation in the parameter space $(W,J,\phi)$, a dimensionless relative deviation $\sigma$ is calculated as
\begin{equation}
\sigma = \frac{{\rm max}\{\Delta(t)\} }{{\rm mean}\{N_{\rm err}(t)\}}, \quad \quad t \in [0, t_{\rm max}].
\label{relative_error}
\end{equation}
where $N_{\rm err}(t)=||\vec{\chi}(t)||$ is the amplitude of the time derivative of the wave function,
\begin{eqnarray}
N_{\rm err}(t) & = & \sqrt{-\langle\frac{\partial}{\partial t}\Psi(t)|\frac{\partial}{\partial t}\Psi(t)\rangle} \nonumber \\
& = & \sqrt{\langle {\rm D}^M_{2}(t)|\hat{H}^2|{\rm D}^M_{2}(t)\rangle}.
\end{eqnarray}

With the wave function $|{\rm D}^M_2(t)\rangle$ obtained, the total energy $E_{\rm total} = E_{\rm ex}+ E_{\rm ph}+ E_{\rm ex-ph}$ is calculated, where $E_{\rm ex} = \langle {\rm D}^M_{2}(t)|\hat{H}_{\rm ex}|{\rm D}^M_{2}(t) \rangle, ~ E_{\rm ph} = \langle {\rm D}^M_{2}(t)|\hat{H}_{\rm ph}|{\rm D}^M_{2}(t) \rangle$, and $E_{\rm ex-ph} = \langle {\rm D}^M_{2}(t)|\hat{H}_{\rm ex-ph}^{\rm o.d.}|{\rm D}^M_{2}(t) \rangle$ (see Eq.~(\ref{energy_equation})).  Additionally, the exciton probability $P_{\rm ex}(t, n)$ and the exciton momentum probability $P_{\rm ex}(t, k)$ are also calculated
\begin{eqnarray}
P_{\rm ex}(t, n) &=& \langle {\rm D}^M_{2}(t) |\hat{a}_{n}^\dag\hat{a}_{n}|{\rm D}^M_{2}(t)\rangle, \nonumber \\
P_{\rm ex}(t, k) &=& \langle {\rm D}^M_{2}(t) |\hat{a}_{k}^\dag\hat{a}_{k}|{\rm D}^M_{2}(t)\rangle,
\label{PX}
\end{eqnarray}
where $\hat{a}_k^{\dag}$ ($\hat{a}_k$) is the creation (annihilation) operator of the exciton with the exciton momentum $k$,
\begin{equation}
\hat{a}_k^{\dag} = N^{-1/2}\sum_n e^{-ikn}\hat{a}_n^{\dag}, \quad \hat{a}_n^{\dag} =  N^{-1/2}\sum_k e^{ikn}\hat{a}_k^{\dag}.
\label{excitonmomentum}
\end{equation}
We then calculate the mean square displacement ${\rm MSD}$ $(t)$ of the exciton probability $P_{\rm ex}(t, n)$ as a function of time $t$,
\begin{eqnarray}
c (t)& = &\sum_{n}^{N}nP_{\rm ex}(t, n), \nonumber \\
{\rm MSD}~(t) & = &	\sum_{n}^{N}\left[n-c(t)\right]^{2}P_{\rm ex}(t, n),
\label{MSD}
\end{eqnarray}
where $c (t)$ describes the centroid motion of the exciton probability. In the exciton momentum representation, the counterpart, $k-{\rm MSD}$ $(t)$ denotes the degree of deviation of the state at the time $t$ from the initial state, as shown in the following,
\begin{eqnarray}
c_k(t)& = &\sum_{k=-\pi}^{\pi}kP_{\rm ex}(t, k), \nonumber \\
k-{\rm MSD}~(t) & = &	\sum_{k=-\pi}^{\pi}\left[k-c_k(t)\right]^{2}P_{\rm ex}(t, k),
\label{MSDk}
\end{eqnarray}
where $c_k(t)$ illustrates the centroid motion of the exciton momentum probability.

\section{Results and discussions}
\label{Numerical results and discussions}

\subsection{The multi-D$_2$ Davydov {\it Ansatz}}
\label{Considering a one-dimensional stack of planar conjugated
molecules}
In this subsection, dynamics of Hamiltonian (\ref{Holstein}) is described fully quantum mechanically using the multi-$\rm D_2$ {\it Ansatz} with sufficiently large multiplicity $M$, yielding numerically accurate quantum dynamics at zero temperature \cite{zhou_15}.

\begin{figure}[tbp]
\centering
\includegraphics[scale=0.4]{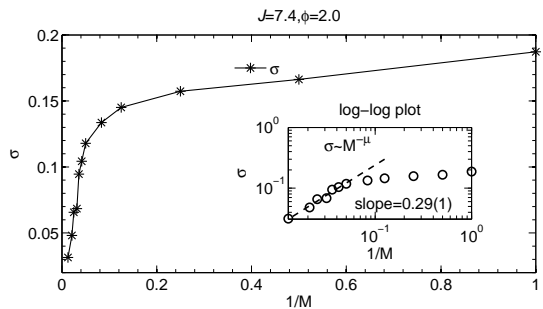}
\caption{The relative deviation $\sigma$ of the multi-${\rm D}_2$ {\it Ansatz} is displayed as a function of $1/M$ for a commonly used set of parameters with off-diagonal coupling $\phi=2.0$ and large transfer integral $J=7.4$. In the inset, the relationship $\sigma \sim M^{\mu}$ is displayed on a log-log scale and the dashed line represents a power-law fit.}
\label{phi_diff_M_J7.4}
\end{figure}
We first test the accuracy of our multi-D${_2}$ {\it Ansatz} with parameters extracted from Refs.~[\citenum{troisi_2006, si_2015}] (this parameter set was extensively used to study realistic models of pentacene molecules). As shown in Fig.~\ref{phi_diff_M_J7.4}, the relative deviation $\sigma$ goes to zero as the multiplicity $M$ approaches infinity. A log-log plot of ($\sigma,1/M$) (inset) indicates a power-law relationship with an exponent of $\mu=0.29 (1)$, further inferring a numerically exact solution in the limit of $M \to \infty$. The largest relative deviation $\sigma$ is found for the single D$_2$ {\it Ansatz}. As presented in Appendix \ref{Fully quantum description of the semiclassical Hamiltonian}, the SSH Hamiltonian is equivalent to the Holstein Hamiltonian with off-diagonal coupling. The equivalence between the semiclassical method and the variational method using the single $\rm D_2$ {\it Ansatz} is shown in Appendix \ref{Comparison between variational method and semi-classical method}. Therefore, this implies that the accuracy of the semi-classical Ehrenfest dynamics can be quantified by the relative deviation of the single D$_2$ {\it Ansatz}. The variational method with sufficiently large $M$ fully takes into account the quantum effects, yielding a much more accurate result than that with the single D$_2$ {\it Ansatz}, which is equivalent to the semi-classical method. For example, $\sigma$ of the D$_2^{M=16}$ {\it Ansatz} in Fig.~\ref{phi_diff_M_J7.4} is smaller than $0.1$, thus the multiplicity of $M=16$ is employed to explore polaron dynamics in following subsections, unless otherwise specified.

\begin{figure}[tbp]
\centering
\includegraphics[scale=0.4]{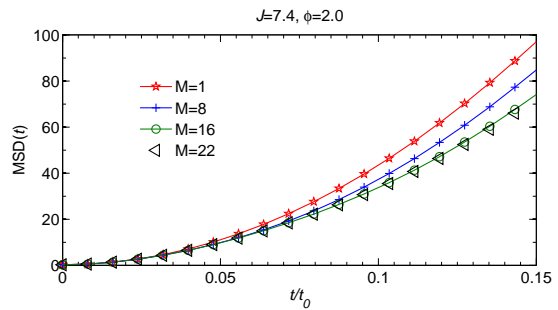}
\caption{${\rm MSD}$ $(t)$ of the exciton for the case of $J=7.4$ and $\phi=2.0$ is obtained from the single $\rm D_2^{M=1}$, the $\rm D_2^{M=8}$, the $\rm D_2^{M=16}$, and the $\rm D_2^{M=22}$ {\it Ansatz}, respectively.}
\label{MSD_J7.4}
\end{figure}

In order to further compare the performance of our variational method using the multi-D$_2$ {\it Ansatz} and that of the semi-classical method, the exciton movement is studied by calculating the mean square displacement ${\rm MSD}$ $(t)$. As shown in Fig.~\ref{MSD_J7.4}, the amplitudes of ${\rm MSD}$ $(t)$ from the fully quantum variational method using the $\rm D_2^{M=8}$, the $\rm D_2^{M=16}$, and the $\rm D_2^{M=22}$ {\it Ansatz} are smaller than that from the semi-classical Ehrenfest method (equivalent to the single $\rm D_2$ {\it Ansatz}), and ${\rm MSD}$ $(t)$ shows apparent convergence as the multiplicity $M$ exceeds $16$. This result is in agreement with that by the IDC approach, which the carrier is found to be less mobile in comparison with that of original Ehrenfest method \cite{yao_2012, dong_2016}. In this case, the transfer integral is much larger than the exciton-phonon coupling and makes more contribution to the movement of the wave front in the carrier propagation. Consequently, the exciton-phonon coupling leads to localization of the wave front. The Ehrenfest method treats the phonons semi-classically and underestimate the confinement effect of the exciton-phonon coupling on the wave function. Therefore the reduction of the mobility is attributed to the quantum mechanical description of the phonons and the electron-phonon coupling. We note, however, the change of ${\rm MSD}$ $(t)$ depends on parameter regimes. In some other cases (e.g., $J=0$ and $\phi=1.0$), phonon assisted transport dominates the exciton movement as discussed in Ref.~[\citenum{zhou_15}].

\subsection{Polaron dynamics in exciton momentum representation}
\label{Different parameter regimes}

\begin{figure}[tbp]
\centering
\includegraphics[scale=0.75]{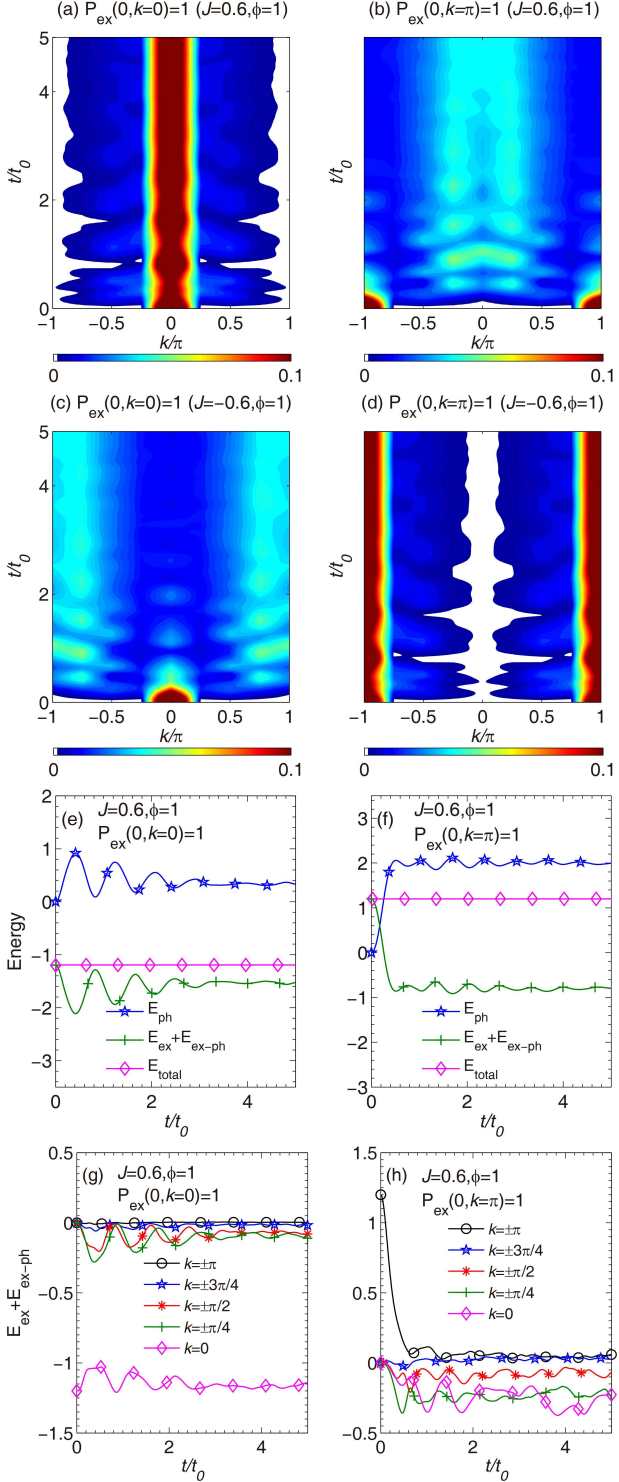}
\caption{ (a)-(d) Time evolution of the exciton momentum probability $P_{\rm ex}(t, k)$ displayed in two columns is obtained using excitonic initial conditions: $P_{\rm ex}(0,k=0)=1$ (left column) and $P_{\rm ex}(0,k=\pi)=1$ (right column). Two transfer integrals, $J=0.6$ and $-0.6$, are used together with the same off-diagonal coupling strength of $\phi=1$, respectively. Energies for the case of $J=0.6$ and $\phi=1$ are plotted for (e) $P_{\rm ex}(0,k=0)=1$ and (f) $P_{\rm ex}(0,k=\pi)=1$. Energies of the exciton and the exciton-phonon coupling are displayed for each exciton momentum $k$ using the initial condition of (g) $P_{\rm ex}(0,k=0)=1$ and (h) $P_{\rm ex}(0,k=\pi)=1$. The number of sites $N=8$ is fixed in these calculations.}
\label{ex_k_initial}
\end{figure}

In this subsection, we explore impacts of off-diagonal coupling on the exciton movement in the exciton momentum representation by using the multi-D$_2$ {\it Ansatz}. Without the exciton-phonon coupling, the Hamiltonian of the bare exciton can be described by the first term of Hamiltonian~(\ref{Holstein}), $\hat H_{\rm ex}$, and the energy band is $E(k)=-2J\cos(k)$. The exciton energy and the exciton momentum are constants of motion. However, in the presence of the exciton-phonon coupling, the exciton momentum may move away from initial values and the exciton energy would be dissipated.

The left and the right columns of Figs.~\ref{ex_k_initial}(a)-(d) present the time evolution of the exciton momentum probability $P_{\rm ex}(t, k)$, starting from initial conditions $P_{\rm ex}(0, k)=1$ at $k=0$ and $\pi$, respectively. $P_{\rm ex}(t, k)$ redistributes toward a quasi stationary state, where no net energy transfer takes place between the exciton and the phonons, as shown in Figs.~\ref{ex_k_initial}(e) and (f). Notwithstanding the difference of initial excitonic conditions, $P_{\rm ex}(t, k)$ in the case of Figs.~\ref{ex_k_initial}(a) and (b) still relaxes to the same stationary regime, where final $P_{\rm ex}(t, k)$ is centered around $k=0$. As for $J=-0.6$, $P_{\rm ex}(t, k)$ is centered around $k=\pm\pi$, as shown in Figs.~\ref{ex_k_initial}(c) and (d). Moreover, the exciton momentum pattern in Fig.~\ref{ex_k_initial}(a) is shifted by $\pi$ comparing to that in Fig.~\ref{ex_k_initial}(d) because the Brillouin zone of the former is $\pi$ shifted from that of the latter, and the shift also occurs for $P_{\rm ex}(t, k)$ in Figs.~\ref{ex_k_initial}(b) and (c).

The energy relaxation process is known to be accompanied with a redistribution of the exciton momentum probability \cite{dorfner_2015}. With regard to the Holstein model with diagonal coupling, the exciton kinetic energy is transferred into the phonons, ending up with a constant value of $E_{\rm ex}+E_{\rm ex-ph}$ \cite{janez_2012}. However, energy relaxation in the Holstein model with off-diagonal coupling is still not well understood. In order to clarify this issue, we consider time evolution of the exciton and the phonon energy. Figs.~\ref{ex_k_initial}(e) and (f) show the time evolution of energies in the case of $J=0.6$ and $\phi=1$ for $P_{\rm ex}(0, k=0)=1$ and $P_{\rm ex}(0, k=\pi)=1$, respectively. Under this parameter set, the initial exciton energy of $P_{\rm ex}(0, k=0)=1$ is the lowest, that of $P_{\rm ex}(0, k=\pi)=1$ is the highest, and those of other initial conditions fall in between. After the transfer integral is changed to $J=-0.6$, due to a phase shift of the Brillouin zone in the exciton momentum space, the initial exciton energy of $P_{\rm ex}(0, k=0)=1$ becomes the maximal while that of $P_{\rm ex}(0, k=\pi)=1$ turns into the minimal for all initial excitonic conditions. As a result, identical energy relaxation processes occur despite that transfer integrals have opposite signs. Thus, only the case of $J=0.6$ and $\phi=1$ is displayed for simplicity. At $t=0$, the phonons are in their vacuum states. Later, the incident exciton wave fronts generate phonons via the exciton-phonon coupling. As a consequence, the exciton energy is transferred to the phonon degrees of freedom. After a fast relaxation process, both the energies of the exciton and the phonons reach steady values. $E_{\rm ex}+E_{\rm ex-ph}$ in the steady state settles around $-2|J|$, which corresponds to the energy minimum of the exciton in the absence of the exciton-phonon coupling. In order to identify the energy contribution of each exciton momentum, we also investigate $E_{\rm ex}+E_{\rm ex-ph}$ in the exciton momentum representation. As plotted in Figs.~\ref{ex_k_initial}(g) and (h), the initial $E_{\rm ex}+E_{\rm ex-ph}$ is $-1.2$ and $1.2$, respectively. After relaxation, the momentum of $k=0$ becomes the main contributor of $E_{\rm ex}+E_{\rm ex-ph}$ for both cases, and also determines the locations of the quasi stationary regime after the exciton momentum redistribution.

\begin{figure}[tbp]
\centering
\includegraphics[scale=0.52]{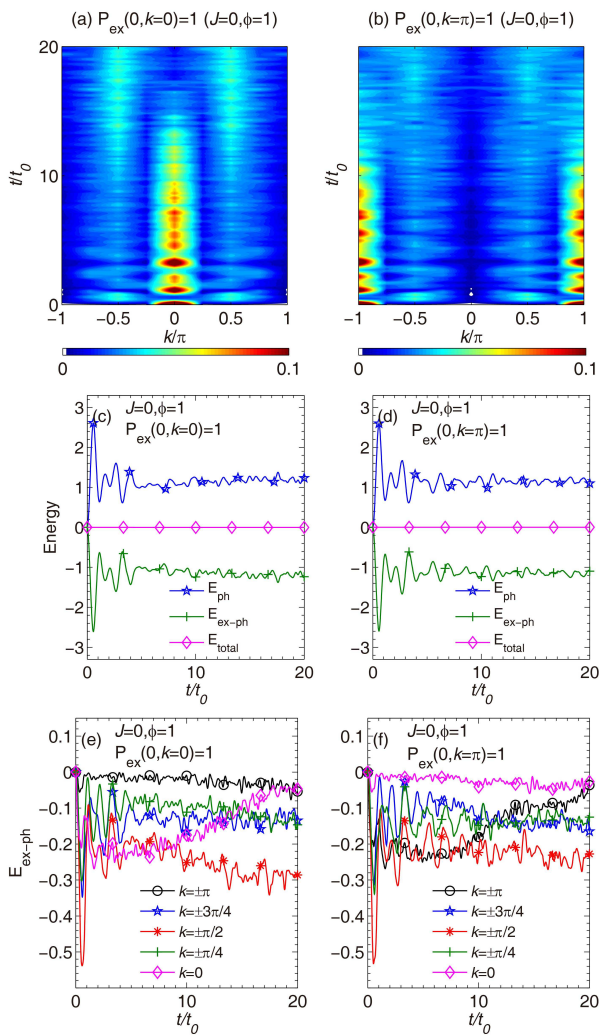}
\caption{Time evolution of the exciton momentum probability $P_{\rm ex}(t, k)$ for $J=0$ and $\phi=1$ is obtained using two initial conditions: (a) $P_{\rm ex}(0, k=0)=1$ (left column) and (b) $P_{\rm ex}(0, k=\pi)=1$ (right column). Corresponding energies are displayed in (c) and (d). The contribution to the exciton-phonon interaction energy from each exciton momentum are shown in (e) and (f).}
\label{ex_k_initial_off}
\end{figure}

Fig.~\ref{ex_k_initial_off} presents the time evolution of the exciton momentum probability in the absence of the transfer integral. We set the initial excitonic conditions $P_{\rm ex}(0,k)=1$ of $k=0$ and $\pi$ in the left ((a),(c) and (e)) and the right ((b),(d) and (f)) column of Fig.~\ref{ex_k_initial_off}, respectively. Akin to the cases of $J=0.6$ and $\phi=1$ in Fig.~\ref{ex_k_initial}, by comparing $P_{\rm ex}(t,k)$ with two types of initial conditions, it is found that the exciton momentum probabilities redistribute and become centered around the same regimes, as shown in Figs.~\ref{ex_k_initial_off}(a) and (b). Even in the absence of the transfer integral, the exciton can still be transported by the off-diagonal coupling. Figs.~\ref{ex_k_initial_off}(c) and (d) plot the time evolution of the phonon energy and the exciton-phonon interaction energy. As also shown in Figs.~\ref{ex_k_initial_off}(c) and (d), for $0<t\leqslant t_0$, the amplitudes of both $E_{\rm ph}$ and $E_{\rm ex-ph}$ reach their peaks and fluctuate until the exciton and the phonons cease to exchange energy at $t=10t_0$. The energy relaxation process only involves $E_{\rm ex-ph}$ because $E_{\rm ex}$ is always zero. As presented in Fig.~\ref{ex_k_initial_off}(e), $E_{\rm ex-ph}$ of each exciton momentum undergoes three stages during the energy relaxation process. During $0<t\leqslant t_0$, they all show strong oscillations with largest amplitudes. At the intermediate stage of $t_0<t\leqslant10t_0$, the energies of $k=\pm\pi/2$ compete with that of $k=0$. For $t>10t_0$, the contribution of the energy of $k=0$ to $E_{\rm ex-ph}$ reduces to almost zero, leaving the energy of $k=\pm\pi/2$ to be the dominant energy contributor. As for the case of the initial condition $P_{\rm ex}(0,k=\pi)=1$ as shown in Fig.~\ref{ex_k_initial_off}(f), the competition at the second stage of $t_0<t\leqslant10t_0$ occurs between the energies of $k=\pm\pi$ and $\pm\pi/2$ instead, and the energy of $k=\pm\pi/2$ also turns out to be the prominent contributor to $E_{\rm ex-ph}$. Consequently, the exciton momentum probability finally becomes centered around $k=\pm\pi/2$ as shown in Figs.~\ref{ex_k_initial_off}(a) and (b).

\subsection{Effect of transfer integral and off-diagonal coupling on exciton transport}
\label{effect of transfer integral and off-diagonal coupling}

\begin{figure}[tbp]
\centering
\includegraphics[scale=0.7]{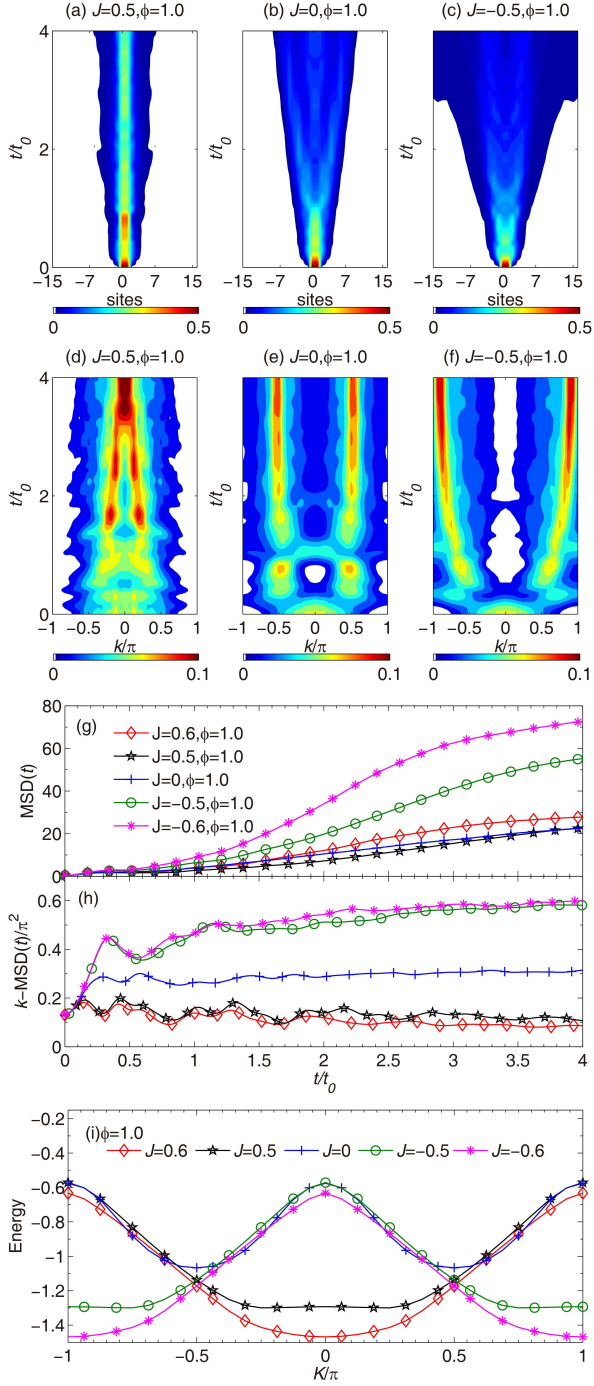}
\caption{Time evolution of the exciton probability $P_{\rm ex}(t, n)$ for the case of the off-diagonal coupling ($\phi=1.0$) is obtained with transfer integrals (a) $J=0.5$, (b) $0$, and (c) $-0.5$; Related time evolution of the exciton momentum probability $P_{\rm ex}(t, k)$ is shown in (d)-(f); (g) ${\rm MSD}~(t)$ of $J=0.6$, $0.5$, $0$, $-0.5$ and $-0.6$ together with $\phi=1.0$ is plotted in the site representation; (h) $k-{\rm MSD}~(t)$ is displayed in the exciton momentum representation; (i) Energy bands of the ground state are obtained from the Toyozawa {\it Ansatz}. Teh number of sites $N=32$ is fixed in these calculations.}
\label{ex_site_k_msd_band}
\end{figure}

In this subsection, we investigate the influence of the transfer integral and the off-diagonal coupling on the exciton transport of Hamiltonian~(\ref{Holstein}).

By tuning the transfer integral, contributions of the transfer integral and the off-diagonal coupling on the exciton movement are examined, as shown in Fig.~\ref{ex_site_k_msd_band}. It can be shown in the site representation that the off-diagonal exciton-phonon coupling plays a crucial role in polaron transport \cite{tamura_2012_85, tamura_2012_86}. As shown in Fig.~\ref{ex_site_k_msd_band}(b), the off-diagonal coupling is the only agent for exciton movement in the absence of the transfer integral, also known as phonon-assisted transport \cite{zhao_1994}. When both the off-diagonal coupling and the direct, phonon free exchange transfer are present, because of the competition between them, the extion transport may be inhibited, as shown in Fig.~\ref{ex_site_k_msd_band}(a). The self-trapping phenomenon is expected due to the competition between the off-diagonal coupling and the transfer integral when the energy bands is flattened at the Brollouin zone center \cite{zh_97}. In this work, the Toyozawa {\it Ansatz} is adopted to study the ground state energy bands of the Holstein model using the variational method. As presented in Fig.~\ref{ex_site_k_msd_band}(i), the lowest energy band of $J=0.5$ and $\phi=1.0$ meets the self-trapping condition, and we can thereby take this case as an example to study the self-trapped exciton from the perspective of dynamics. In agreement with our expectation, $P_{\rm ex}(t, n)$ turns out to be localized in Fig.~\ref{ex_site_k_msd_band}(a). By directly flipping the sign of the transfer integral to $J=-0.5$, the exciton wave fronts are found to move considerably, as shown in Fig.~\ref{ex_site_k_msd_band}(c). Via ${\rm MSD}~(t)$ as defined in Eq.~(\ref{MSD}), the expansion of the exciton wave packets is further investigated for $J=-0.6$, $-0.5$, $0$, $0.5$ and $0.6$. As plotted in Fig.~\ref{ex_site_k_msd_band}(g), the amplitude of ${\rm MSD}~(t)$ for $J=0$ and $\phi=1$ is smaller than those of other cases with non-zero transfer integrals, except the self-trapped case of $J=0.5$ and $\phi=1.0$.

In the crystal momentum representation, the underlying physics of the ground states can be elucidated, where the crystal momentum is denoted as $K$ (see Eq.~(\ref{total momentum})). The Toyozawa {\it Ansatz} is a time independent translationally invariant trial state, viewed as a superposition of the replicas of the D$_2$ {\it Ansatz} displaced to every lattice site, weighed by a phase factor of the total momentum \cite{zh_97}. We analyzed the energy bands of the ground states obtained from the Toyozawa {\it Ansatz} (see Appendix~\ref{The Toyozawa Ansatz}).
In the the off-diagonal coupling only case ($J=0$), the minima of the band are located at $K=\pm\pi/2$. The addition of positive (negative) transfer integrals moves the minima towards the center (boundary). In particular, as mentioned above, the case of $J=0.5$ flattens the band at the center of the Brillouin zone, leading to the largest effective mass of all studied cases, in accord with the self-trapping of $P_{\rm ex}(t, n)$ in Fig.~\ref{ex_site_k_msd_band}(a).

The effect of the transfer integrals on the exciton movement in the presence of the off-diagonal coupling is further examined in the exciton momentum representation in Figs.~\ref{ex_site_k_msd_band}(d)-(f) and (h). The exciton is created in the profile of $\left(2+\cos k\right)/2N$ in the momentum space as we excite two nearest neighboring sites initially (see Appendix \ref{Equations}). In the subsequent relaxation process, $P_{\rm ex}(t, k)$ redistributes and becomes localized in a quasi stationary region, and the mean square displacement of the exciton momentum $k-{\rm MSD}~(t)$ approaches a plateau, as shown in Fig.~\ref{ex_site_k_msd_band}(h). After the relaxation process, the final $P_{\rm ex}(t, k)$ is found to be determined by a combination of the transfer integral and the off-diagonal coupling strength. For the off-diagonal coupling only case, $P_{\rm ex}(t, k)$ progressively becomes localized around $k=\pm\pi/2$ (Fig.~\ref{ex_site_k_msd_band}(e)). In the case of $J=0.5$ and $0.6$, $P_{\rm ex}(t, k)$ aggregates toward $k=0$, as seen in Fig.~\ref{ex_site_k_msd_band}(d). Similarly, $P_{\rm ex}(t, k)$ of both $J=-0.5$ and $-0.6$ correspond to $\pm\pi$ in Fig.~\ref{ex_site_k_msd_band}(f). In addition, $k-{\rm MSD}~(t)$ for the extreme cases of $P_{\rm ex}(t, k)=\delta_{k,0}$ and $\delta_{k,\pm\pi}$ are $0$ and $2\pi^2$, respectively. As shown in Fig.~\ref{ex_site_k_msd_band}(h), $k-{\rm MSD}~(t)$ of $J=0.6$ is closer to the analytical value of $0$ than that of $J=0.5$, indicating that $P_{\rm ex}(t, k)$ of $J=0.6$ is more localized around the zone center than that of $J=0.5$. Likewise, $k-{\rm MSD}~(t)$ of $J=-0.6$ is nearer to the limited value of $2\pi^2$ than that of $J=-0.5$, illustrating that $P_{\rm ex}(t, k)$ of the former case is more localized around $k=\pm\pi$.

\begin{figure}[tbp]
\centering
\includegraphics[scale=0.45]{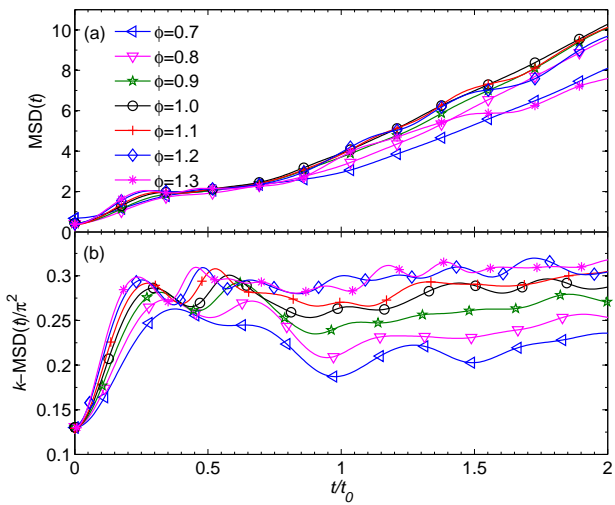}
\caption{(a) ${\rm MSD}~(t)$ of the exciton in the site representation is shown for $J=0$ and $\phi = 0.7, 0.8, 0.9, 1.0, 1.1, 1.2$ and $1.3$; Related $k-{\rm MSD}~(t)$ in the exciton momentum representation is displayed in (b).}
\label{difphi}
\end{figure}

In the site representation, the off-diagonal coupling is known to play the role of assisting the transport of the exciton \cite{tamura_2012_86, zhao_1994}. In Fig.~\ref{difphi}(a), in the absence of transfer integral ($J=0$), we study the dependence of ${\rm MSD}~(t)$ on the off-diagonal coupling strength. It is found that the exciton propagation is facilitated by the off-diagonal coupling, as shown by the site-space ${\rm MSD}~(t)$ in Fig.~\ref{difphi}(a). However, the off-diagonal coupling can be simultaneously an agent for exciton localization. The localization effect of $\phi$ gradually increases with the coupling strength $\phi$ if $\phi$ is greater than a critical value $\phi_c$ \cite{zhao_12}. As shown in Fig.~\ref{difphi}(a), the amplitude of ${\rm MSD}~(t)$ decreases with the off-diagonal coupling strength $\phi$ for $\phi>\phi_c=1.0$.

In the exciton momentum representation, $P_{\rm ex}(t, k)$ all ends up around $k=\pm \pi/2$ for a variety of off-diagonal coupling strengths, and the corresponding $k-{\rm MSD}~(t)$ approaches the same narrow regime around $0.25\pi^2$, which is the theoretical value of $k-{\rm MSD}~(t)$ for $P_{\rm ex}(t, k)=(\delta_{k,\pi/2}+\delta_{k,-\pi/2})/2$, as shown in Fig.~\ref{difphi}(b). However, the relaxation time diverges due to the variance of the off-diagonal coupling. The time for the exciton momentum to reach the stationary regime is inversely related to the off-diagonal coupling strength, because the first stage of the time evolution (t$<0.5$t$_0$) is accompanied by the fast exciton movement in the case of large off-diagonal coupling as presented in Fig.~\ref{difphi}. In addition, the energy bands of various off-diagonal coupling strengths imply that the band width and effective mass are largest for $\phi_c^{\rm static}=1.4$ and get smaller as $\phi$ moves away from $\phi_c^{\rm static}$ \cite{zhao_12}. The localization feature is found both in static and dynamic calculations although the value of $\phi_c$ differs slightly. The off-diagonal exciton-phonon coupling leads to exciton energy dissipation and redistribution of exciton momentum in three typical scenarios corresponding to completely distinguishable band structures (this conclusion is independent of the system size), which may be formed due to a variety of compositions and geometrical structures of the organic materials, defects, doping mechanisms and deformations of CPs \cite{bredas_1984, kuklja_2007}.

\subsection{Temperature effects}
\label{Effect of temperature}
In this subsection, we extend the work to study the effect of finite temperatures on polaron dynamics. The conductivity of polymers has been measured by many workers as a function of temperature \cite{heeger_1988, ishiguro_1992, chen_1994}. The temperature effects have been in contention from a theoretical point of view. For example, Cruzeiro {\it et al.}~claims that Davydov soliton is stable at $T=310$ K \cite{cruz_1988}. Later, a quantum Monte Carlo treatment has shown that the Davydov soliton is unstable above $7$ K \cite{wang_1989}. In this work, several approaches are used to study the temperature effects: a variational method with importance sampling (see Appendix \ref{Initial displacements based on the Boltzmann distributions}), the hierarchical equations of motion (HEOM) method \cite{Tanimura1, ch_15}, and the averaged Hamiltonian method (see Appendix \ref{Thermally averaged Hamiltonian}). The variational method with importance sampling developed by Wang {\it et al.}~simulates thermal fluctuation of phonon modes by sampling the initial phonon displacements based on the Bose distribution, and thus it can deal with Holstein polaron dynamics at both low and high temperatures \cite{wanglu_un}. The HEOM method is numerically exact and is capable to treat any finite temperature, serving as a benchmark here. However, the HEOM method is also numerically expensive and thus impractical when the system size is large. The variational approach can treat large systems once a proper trial wave function is adopted. In order to compare to previous attempts in the literature, the averaged Hamiltonian method has also been used, and we found that this approach is not even suitable for the spin-boson model ({\it i.e.}., \ N=$2$) as shown in Appendix \ref{Different methods for the temperature effect}.

\begin{figure}[tbp]
\centering
\includegraphics[scale=0.43]{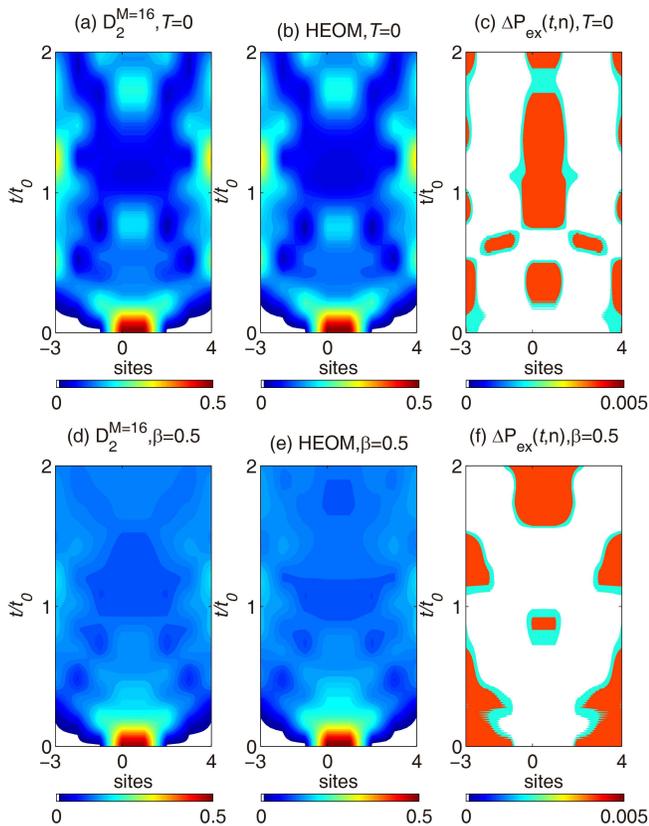}
\caption{Time evolution of the exciton probability $P_{\rm ex}(t, n)$ obtained at $T=0$ and $2/k_B$. $P_{\rm ex}(t, n)$ at $T=0$ obtained from (a) the $\rm D_2^{M=16}$ {\it Ansatz}, (b) the HEOM method, and (c) $\Delta P_{\rm ex}(t, n)$ between the $\rm D_2^{M=16}$ {\it Ansatz} and the HEOM method; $P_{\rm ex}(t, n)$ obtained from (d) the $\rm D_2^{M=16}$ {\it Ansatz}, (e) the HEOM method, and (f) the related $\Delta P_{\rm ex}(t, n)$ at $T=2/k_B$ ($\beta=0.5$).}
\label{ex_temperature}
\end{figure}

Fig.~\ref{ex_temperature} shows polaron dynamics calculated by the multi-D$_2$ {\it Ansatz} with importance sampling and the HEOM method for two temperatures. The calculations are carried out for $J=0.8$ and $\phi=0.3$ in a ring of $N=8$ sites. $P_{\rm ex}(t, n)$ outputs obtained from the D$_2^{M=16}$ {\it Ansatz} and the HEOM method  at $T=0$ are shown and compared in Figs.~\ref{ex_temperature}(a) and (b), respectively. As revealed in Fig.~\ref{ex_temperature}(c), $\Delta P_{\rm ex}(t, n)$, {\it i.e.}, the difference between the two methods, are two orders of magnitude smaller than the value of $P_{\rm ex}(t, n)$, indicating that variational method can be numerically exact at low temperatures with sufficient multiplicity $M$ of the multi-$\rm D_2$ {\it Ansatz}. The phonon displacement $\lambda_{i,q}(t=0)$ is set to zero at $T=0$, while importance sampling is used at $T=2/k_B$ ($\beta=0.5$) to simulate the finite temperature effects with the result displayed in Fig.~\ref{ex_temperature}(d). Similarly, as shown in Fig.~\ref{ex_temperature}(f), differences between results from the two methods are two orders of magnitude smaller than the value of $P_{\rm ex}(t, n)$ in Figs.~\ref{ex_temperature}(d) end (e), inferring that the variational method with importance sampling provides numerically exact results at high temperatures with sufficiently large $M$.

Next, we investigate the influence of thermal fluctuations on exciton transport. For both low and high temperatures, the exciton wave fronts depart from the site of exciton creation and propagate in opposing directions until they meet at the opposite side of the ring. During the time evolution, distinct features observed at zero temperature (Figs.~\ref{ex_temperature}(a) and (b)) are now significantly smeared due to the thermal fluctuations (Figs.~\ref{ex_temperature}(d) and (e)). As shown in Figs.~\ref{ex_temperature}(d) and (e), during $0.2t_0<t<t_0$ the exciton probability is more centered around the site of creation than those of Figs.~\ref{ex_temperature}(a) and (b). For $t\geqslant t_0$, the bright spots shown in Figs.~\ref{ex_temperature}(a) and (b) are significantly quenched in Figs.~\ref{ex_temperature}(d) and (e). These results indicate that the exciton transport is weakened when the temperature is increased, in line with Refs.~[~\citenum{troisi_2006, chang_2011}].

\section{Conclusion}
\label{Conclusions}
In this work, we have studied the dynamics of the Holstein molecular crystal model with off-diagonal coupling using the Dirac-Frenkel time-dependent variational principle and the novel multi-D$_2$ {\it Ansatz}, which is a linear combination of the usual Davydov $\rm D_2$ trial state from the soliton literature. Traditionally used to simulate such dynamics is the semi-classical Ehrenfest method, which has been shown to be equivalent to our time-dependent variational method with the single D$_2$ {\it Ansatz}. Calculation of the relative deviation, which quantifies the {\it Anstaz} accuracy, demonstrates that the variational method with the multi-D$_2$ {\it Ansatz} presents much more accurate results than the semi-classical Ehrenfest dynamics. With a sufficiently large multiplicity, our variational method using the multi-D$_2$ {\it Ansatz} can offer numerically exact solutions. We further compare $\rm MSD$ $(t)$ obtained from the semi-classical method and our variational method with that from the multi-D$_2$ {\it Ansatz}, and find that the mobility is overestimated by the semi-classical method. These results indicate that the description beyond the semi-classical method is essential to quantitatively capture the dynamics of the SSH model.

Secondly, we have explored the underlying physics from the accurate dynamics data for the Holstein model with the off-diagonal coupling. The energy and the momentum of the bare exciton are constants of motion. However, in the presence of the exciton-phonon coupling, the exciton momentum probability is found to redistribute and become centered in stationary regions. We reveal that the momentum redistribution is only determined by the combination of the transfer integral and the off-diagonal strength, and is independent of the initial excitonic conditions used. In addition, in order to study the competition between the transfer integral and the off-diagonal coupling, we investigate the exciton transport within the exciton site and the exciton momentum representation, and the crystal momentum representation. The results show that the combination of the transfer integral and the off-diagonal coupling do not necessarily play a role in enhancing the exciton transport. Moreover, the off-diagonal coupling is demonstrated to be the simultaneous agent of transport and localization in dynamical calculations.

Lastly, the temperature effects are studied using the variational method with importance sampling by employing the multi-D$_2$ {\it Ansatz}. In both the low and high temperature regimes, the time evolution of the exciton probability calculated from the variational method with importance sampling agrees well with that from the numerically exact HEOM method, and can be obtained much more efficiently. The results at the finite temperatures show that fast delocalization of the exciton wave is quenched due to the thermal fluctuations, indicating the weakening of the exciton transport by increasing the temperature.

\section*{Acknowledgments}
We thank Yuta Fujihashi and Zheng Fulu for helpful discussion. Support from the Singapore National Research Foundation through the Competitive Research Programme (CRP) under Project No.~NRF-CRP5-2009-04 is gratefully acknowledged.

\appendix
\section{Fully quantum description of the semiclassical Hamiltonian}
\label{Fully quantum description of the semiclassical Hamiltonian}
The semi-classical Hamiltonian is composed of the electronic and the phonon part $H=H_{el}+H_{ph}$, the electronic part is
\begin{eqnarray}
H_{el}	=	\sum_{n}\left[-J+\alpha\left(u_{n+1}-u_{n}\right)\right]\left(\hat{a}_{n}^{\dagger}\hat{a}_{n+1}+H.c.\right),
\end{eqnarray}
where $J$  is the transfer integral, $\alpha$ the electron-phonon coupling constant, $u_{n}$ the displacement of phonon on the $n$-th site, and $\hat{a}{}_{n}^{\dagger}\left(\hat{a}_{n}\right)$ the creation (annihilation) operators of electron.
The phonon part is
\begin{eqnarray}
H_{ph}	=	\frac{K}{2}\sum_{n}\left(u_{n+1}-u_{n}\right)^{2}+\frac{M}{2}\sum_{n}\dot{u}_{n}^{2},
\end{eqnarray}
in which, $K$ denotes the force constant originating from the $\sigma$ bond between carbon atoms, $K=M\omega_0^{2}$ and $M$ the total mass of a CH-unit for trans-polyacetylene. The combination of the two parts above is identical to  Su-Schrieffer-Heeger (SSH) model used for conductive polymers \cite{heeger_1988}.

Using the quantum mechanical creation and annihilation operators to describe the displacement of the phonon bath,
\begin{eqnarray}
&&u_{n}=\sqrt{\frac{1}{2M\omega}}\left(b_{n}^{\dagger}+b_{n}\right),\nonumber\\
&&\gamma=\sqrt{\frac{1}{2M\omega}}\alpha
\end{eqnarray}
we get $\hat{H}=\hat{H}_{ex}+\hat{H}_{ph}+\hat{H}_{int}$,
with the electronic part
\begin{eqnarray}
\hat{H}_{ex}	=	-J\sum_{n}\left(\hat{a}_{n}^{\dagger}\hat{a}_{n+1}+H.c.\right),
\end{eqnarray}
the phonon part
\begin{eqnarray}
\hat{H}_{ph}	=	\sum_{n}\omega_{0}\hat{b}_{n}^{\dagger}\hat{b}{}_{n},
\end{eqnarray}
and the electron-phonon interaction part
\begin{eqnarray}
&&\hat{H}_{int}	= \gamma\sum_{n,l}\left[\hat{a}_{n}^{\dagger}\hat{a}_{n+1}\left(\hat{b}_{l}+\hat{b}_{l}^{\dagger}\right)\left(\delta_{n+1,l}-\delta_{n,l}\right)\right.\nonumber\\
&&		+\left.\hat{a}_{n}^{\dagger}\hat{a}_{n-1}\left(\hat{b}_{l}+\hat{b}_{l}^{\dagger}\right)\left(\delta_{n,l}-\delta_{n-1,l}\right)\right].
\end{eqnarray}

Fourier transforming the phonon operators into momentum space,
\begin{eqnarray}
&&\hat{b}_{n}^{\dagger}=N^{-1/2}\sum_{q}e^{-iqn}\hat{b}_{q}^{\dagger},\nonumber\\
&&\hat{b}_{n}=N^{-1/2}\sum_{q}e^{iqn}\hat{b}_{q},
\end{eqnarray}
we get
\begin{eqnarray}
&&\hat{H}_{ex} = -J\sum_{n}\left(\hat{a}_{n}^{\dagger}\hat{a}_{n+1}+H.c.\right),\nonumber\\
&&\hat{H}_{ph} = \omega_{0}\sum_{q}\hat{b}_{q}^{\dagger}\hat{b}_{q},\nonumber\\
&&\hat{H}_{ex-ph}^{o.d.} = \gamma N^{-1/2}\sum_{n,q}\left\{\hat{a}_{n}^{\dagger}\hat{a}_{n+1}\left[e^{iqn}\left(e^{iq}-1\right)\hat{b}_{q}+H.c.\right]\right.\nonumber\\
&&		+\left.\hat{a}_{n}^{\dagger}\hat{a}_{n-1}\left[e^{iqn}\left(1-e^{-iq}\right)\hat{b}_{q}+H.c.\right]\right\},
\end{eqnarray}
just being the off-diagonal Holstein polaron model.

\section{Comparison between the variational method using the single $D_2$ {\it Ansatz} and the semi-classical method}
\label{Comparison between variational method and semi-classical method}
In this part, it is shown that the dynamics obtained from the semi-classical method and the variational method using only the single $D_2$ {\it Ansatz} are equivalent for the spin-boson model ($N=2$).
\subsection{The variational method}
The Hamiltonian of the spin-boson model is
\begin{eqnarray}
\hat{H}=\frac{\epsilon}{2}\sigma_{z}+V\sigma_{x}+\omega\hat{a}^{\dagger}\hat{a}+\frac{\lambda}{2}\sigma_{z}(\hat{a}^{\dagger}+\hat{a})+\frac{\phi}{2}\sigma_{x}(\hat{a}^{\dagger}+\hat{a}),\nonumber\\
\label{SBH}
\end{eqnarray}
where $\epsilon$ and $V$ are the spin bias and the tunneling constant, respectively. $\lambda$ ($\phi$) is the diagonal(off-diagonal) coupling strength. $\sigma_x$ and $\sigma_z$ are THE Pauli matrices, $\hat{a}^{\dagger}(\hat{a})$ is the boson creation (annihilation) operator for the phonon of frequency $\omega_0$.

Using the variational principle and the $D_{2}$ {\it Ansatz},
$\left|D_{2}\right\rangle \nonumber\\
=A(t)\left|+\right\rangle e^{\left[f(t)\hat{a}^{\dagger}-f^{*}(t)\hat{a}\right]}\left|0\right\rangle +B(t)\left|-\right\rangle e^{\left[f(t)\hat{a}^{\dagger}-f^{*}(t)\hat{a}\right]}\left|0\right\rangle$
the equations of motion can be obtained,
\begin{eqnarray}
&&0	=	i\dot{A}-\frac{\epsilon}{2}A-VB-\frac{\lambda}{2}A(f+f^{*})-\frac{\phi}{2}B\left(f^{*}+f\right), \nonumber \\
&&0	=	i\dot{B}+\frac{\epsilon}{2}B-VA+\frac{\lambda}{2}B(f+f^{*})-\frac{\phi}{2}A\left(f^{*}+f\right), \nonumber \\
&&0	=i\dot{f}-\omega f-\frac{\lambda}{2}(\left|A\right|^{2}-\left|B\right|^{2})-\frac{\phi}{2}\left(A^{*}B+AB^{*}\right).\nonumber \\
\label{equations of variational method}
\end{eqnarray}

\subsection{The semi-classical method}
The semi-classical Hamiltonian can also be written as
\begin{eqnarray}
\hat{H}=\frac{\epsilon}{2}\sigma_{z}+V\sigma_{x}+\frac{p^{2}}{2m}+\frac{1}{2}m\omega^{2}x^{2}\nonumber \\+\lambda\sqrt{\frac{m\omega}{2}}\sigma_{z}x+\phi\sqrt{\frac{m\omega}{2}}\sigma_{x}x,
\end{eqnarray}
the electronic state is described by the wave function $\left|\psi\right\rangle =A\left(t\right)\left|+\right\rangle +B\left(t\right)\left|-\right\rangle$, $m$ is the effective mass of the phonon.
The equations of motion from the semi-classical formalism are
\begin{eqnarray}
i\dot{A}	&=&	\left(\frac{\epsilon}{2}+\sqrt{\frac{m\omega}{2}}\lambda x\right)A+VB+\phi\sqrt{\frac{m\omega}{2}}xB,\nonumber\\
i\dot{B}	&=&	-\left(\frac{\epsilon}{2}+\sqrt{\frac{m\omega}{2}}\lambda x\right)B+VA+\phi\sqrt{\frac{m\omega}{2}}xA,\nonumber\\
\dot{x} &=& v,\nonumber\\
\dot{v} &=& -\omega^{2}x-\lambda\sqrt{\frac{\omega}{2m}}\left(\left|A\right|^{2}-\left|B\right|^{2}\right)\nonumber \\
&&-\phi\sqrt{\frac{\omega}{2m}}\left(A^{*}B+AB^{*}\right).
\label{equations of semi-classical method}
\end{eqnarray}

\subsection{Comparison}
We now compare the equations of motion from the variational method and the semi-classical method. From Eq.~(A$7$) of Ref.~[\citenum{fran_2016}], we get $f	=	\frac{1}{\sqrt{2m\omega}}\left(m\omega x+ip\right)$ and $f+f^{*}=\sqrt{{2m\omega}}x$. After we put this into the last equation of Eq.~(\ref{equations of variational method}), we get
\begin{eqnarray}
&&0	=	i\frac{1}{\sqrt{2m\omega}}\left(m\omega\dot{x}+i\dot{p}\right)-\omega\frac{1}{\sqrt{2m\omega}}\left(m\omega x+ip\right)\nonumber\\
&&		-\frac{\lambda}{2}(\left|A\right|^{2}-\left|B\right|^{2})-\frac{\phi}{2}\left(A^{*}B+AB^{*}\right)
\label{compared}
\end{eqnarray}
The real part of Eq.~(\ref{compared}) agrees with the fourth equation of Eq.~(\ref{equations of semi-classical method}),
and imaginary part of Eq.~(\ref{compared}) is equal to the third equation of Eq.~(\ref{equations of semi-classical method}), proving the equivalence of the semi-classical method and variational method using only the single $D_2$ {\it Ansatz}.

In conclusion, the expectation value of position $x$ and momentum $p$ obtianed from the semi-classical method agree with the ones from the variational method using THE single $D_{2}$ {\it Ansatz}.

\section{The multi-${\rm D}_2$ trial states}
\label{Equations}
The energies of the system are given by the following equations,
\begin{eqnarray}
&&\left\langle D_{2}^M\left(t\right)\right|H_{ex}\left|D_{2}^M\left(t\right)\right\rangle=-J\sum_{i,j}^{M}\sum_{n}\psi_{jn}^{\ast}\left(\psi_{i,n+1}+\psi_{i,n-1}\right)S_{ji},\nonumber \\
&&\left\langle D_{2}^M\left(t\right)\right|H_{ph}\left|D_{2}^M\left(t\right)\right\rangle=\omega_{0}\sum_{i,j}^{M}\sum_{n}\psi_{jn}^{\ast}\psi_{in}\sum_{q}\lambda_{jq}^{\ast}\lambda_{iq}S_{ji},\nonumber \\
&&\left\langle {\rm D}^M_{2}\left(t\right)\right|H_{ex-ph}^{o.d.}\left|{\rm D}^M_{2}\left(t\right)\right\rangle=\frac{1}{2}N^{-1/2}\phi\sum_{n,q}\sum_{i,j}^{M}\omega_{q}S_{ji}\nonumber \\
&& \times\{\psi_{jn}^{\ast}\psi_{i,n+1}[e^{iqn}(e^{iq}-1)\lambda_{iq}+e^{-iqn}(e^{-iq}-1)\lambda_{jq}^{\ast}]\nonumber \\
&& +\psi_{jn}^{\ast}\psi_{i,n-1}[e^{iqn}(1-e^{-iq})\lambda_{iq}+e^{-iqn}(1-e^{iq})\lambda_{jq}^{\ast}]\},
\label{energy_equation}
\end{eqnarray}
where the Debye-Waller factor is formulated as
\begin{eqnarray}
S_{ij} & = & \left\langle \lambda_{i}\vert\lambda_{j}\right\rangle, \nonumber \\
& = & \exp\left\{\sum_{q}\lambda_{jq}^{\ast}\lambda_{iq}-\frac{1}{2}(|\lambda_{iq}|^2+|\lambda_{jq}|^2)\right\}.
\label{A2}
\end{eqnarray}
In addition, the energies can be converted to the exciton momentum representation by using
\begin{eqnarray}
&&\psi_{in}=N^{-1/2}\sum_{k}e^{-ikn}\psi_{ik},\nonumber\\
&&\psi_{in}^{\ast}=N^{-1/2}\sum_{k}e^{ikn}\psi_{ik}^{\ast}.
\end{eqnarray}

The Dirac-Frenkel variational principle results in equations of motion including
\begin{eqnarray}
&&		-i\sum_{i}\dot{\psi}_{in}S_{ki}\nonumber \\
	=	 &&+\frac{i}{2}\sum_{i}\psi_{in}\sum_{q}\left(2\lambda_{kq}^{\ast}\dot{\lambda}_{iq}-\dot{\lambda}_{iq}\lambda_{iq}^{\ast}-\lambda_{iq}\dot{\lambda}_{iq}^{\ast}\right)S_{k,i}\nonumber \\
&&		+J\sum_{i}\left(\psi_{i,n+1}+\psi_{i,n-1}\right)S_{ki}\nonumber \\
&&		-\omega_{0}\sum_{i}\psi_{in}\sum_{q}\lambda_{kq}^{\ast}\lambda_{iq}S_{ki}\nonumber \\
&&		-\frac{1}{2}N^{-1/2}\phi\sum_{i}\sum_{q}\omega_{q}\psi_{i,n-1}[e^{iqn}(1-e^{-iq})\lambda_{iq}\nonumber \\
&&+e^{-iqn}(1-e^{iq})\lambda_{kq}^{\ast}]S_{ki}, \nonumber \\
\end{eqnarray}
and
\begin{eqnarray}
&&		-i\sum_{i}\sum_{n}\psi_{kn}^{\ast}\dot{\psi}_{in}\lambda_{iq}S_{ki}-i\sum_{i}\sum_{n}\psi_{kn}^{\ast}\psi_{in}\dot{\lambda}_{iq}S_{ki}\nonumber \\
&&-\frac{i}{2}\sum_{i}\sum_{n}\psi_{kn}^{\ast}\psi_{in}\lambda_{iq}S_{k,i}\nonumber\\
&&\sum_{p}\left(2\lambda_{kp}^{\ast}\dot{\lambda}_{ip}-\dot{\lambda}_{ip}\lambda_{ip}^{\ast}-\lambda_{ip}\dot{\lambda}_{ip}^{\ast}\right)\nonumber \\=
&&		J\sum_{i}\sum_{n}\psi_{kn}^{\ast}\left(\psi_{i,n+1}+\psi_{i,n-1}\right)\lambda_{iq}S_{k,i}\nonumber \\
&&		-\sum_{i}\sum_{n}\psi_{kn}^{\ast}\psi_{in}\left(\omega_{0}+\omega_{0}\sum_{p}\lambda_{kp}^{\ast}\lambda_{ip}\right)\lambda_{iq}S_{ki}\nonumber \\
&&-\frac{1}{2}N^{-1/2}\phi\sum_{n}\sum_{i}\omega_{q}\psi_{kn}^{\ast}[\psi_{i,n+1}e^{-iqn}(e^{-iq}-1)\nonumber\\
&&+\psi_{i,n-1}e^{-iqn}(1-e^{iq})]S_{ki}\nonumber \\
&&-\frac{1}{2}N^{-1/2}\phi\sum_{n}\sum_{i}\left(\psi_{k,n+1}^{\ast}\psi_{i,n}+\psi_{kn}^{\ast}\psi_{i,n+1}\right)\lambda_{iq}\nonumber\\
&&\sum_{p}\omega_{p}[e^{ipn}(e^{ip}-1)\lambda_{ip}+e^{-ipn}(e^{-ip}-1)\lambda_{kp}^{\ast}]S_{k,i}.\nonumber\\
\end{eqnarray}\

It should be noted that the main results of this work are calculated from the above equations of motion. The equations of motion are solved numerically by means of the fourth-order Runge-Kutta method. The exciton initially sits on two nearest-neighboring sites, {\it i.e}\, $\psi_n=(\delta_{n,N/2}+\delta_{n,N/2+1})/\sqrt2$. At $T=0$, the phonon state is at the vacuum state, {\it i.e}\, $\lambda_{iq}=0$. In order to avoid singularity, the uniformly distributed noise $[-10^{-5},10^{-5}]$ is added to the initial variational parameters $\psi_{in}$ and $\lambda_{iq}$ at $t=0$. More than one hundred samples are averaged to get rid of the influence of the noise and reach convergent results in the simulations.

\section{The Toyozawa Ansatz}
\label{The Toyozawa Ansatz}
Our interest here includes the polaron ground-state energy band, computed as
\begin{eqnarray}
E(K)=\langle \Psi(K) |\hat{H}|\Psi(K) \rangle,
\end{eqnarray}
where$|\Psi(K) \rangle$ is an appropriately normalized, delocalized trial state, and $\hat{H}$ is the system Hamiltonian. The joint crystal
momentum is indicated by the $K$. It should be noted that the crystal momentum operator commutes with the system Hamiltonian, and energy eigenstates are also eigenfunctions of the crystal momentum. Therefore, variations for distinct $K$ are independent. The set of $E(K)$ constitutes a variational estimate (an upper bound) for the polaron energy band. The relaxation iteration technique, viewed as an efficient method for identifying energy minima of a complex variational system, is adopted to obtain numerical solutions to a set of self-consistency equations derived from the
variational principle. To achieve efficient and stable iterations toward the variational ground state, one may take advantage of the continuity of the ground state with respect to small changes in system parameters over most of the phase diagram and may initialize the iteration using a reliable ground state already determined at some nearby points in parameter space. Starting from those limits where exact solutions can be obtained analytically and executing a sequence
of variations along well-chosen paths through the parameter space using solutions from one step to initialize the next, the whole parameter space can be explored.

As the D$_2$ {\it Ansatz} is a localized state from the soliton theory. It can be delocalized into the Toyozawa {\it Ansatz}, which is the Bloch state with the designated crystal momentum, via a projection operator $\hat P_K$.
\begin{eqnarray}
\hat P_K = N^{-1}\underset{n}{\sum }e^{i(K-\hat P) n}=\delta(K-\hat P),
\label{total momentum}
\end{eqnarray}
where
\begin{eqnarray}
\hat P= \underset{k}{\sum }k a_{k}^{\dagger }a_{k}+\underset{q}{\sum }q b_{q}^{\dagger }b_{q},
\end{eqnarray}

After the delocalization onto the usual $\rm D_2$ {\it Ansatz}, the Toyozawa {\it Ansatz} is given by
\begin{eqnarray}
|\Psi_2(K')\rangle =|K' \rangle \langle K'|K' \rangle^{-1/2},
\end{eqnarray}

\begin{eqnarray}
|K' \rangle  & = & \underset{n}{\sum }e^{iK' n}\underset{n1}{\sum }\psi _{n_1-n}^{K' }a_{n_1}^{\dagger } \\
&&\exp [-\underset{n2}{\sum }(\lambda
_{n_2-n}^{K' }b_{n_2}^{\dagger }-{\rm H.c.})]|0\rangle, \nonumber
\end{eqnarray}
where $\psi _{n_1-n}^{K' }$ is the exciton amplitude and $\lambda_{n_2-n}^{K' }$ is the phonon displacement.

\section{Alternative approaches to temperature effects}
\label{Different methods for the temperature effect}

We aim to investigate the effect of the temperature by comparing following approaches: the averaged Hamiltonian (see Appendix \ref{Thermally averaged Hamiltonian}), the variational method with importance sampling (see Appendix \ref{Initial displacements based on the Boltzmann distributions}), and the numerically exact HEOM method. The spin-boson model, {\it i.e}, \ a Holstein model with N$=2$, is taken as the simplest example.

\begin{figure}[tbp]
\centering
\includegraphics[scale=0.4]{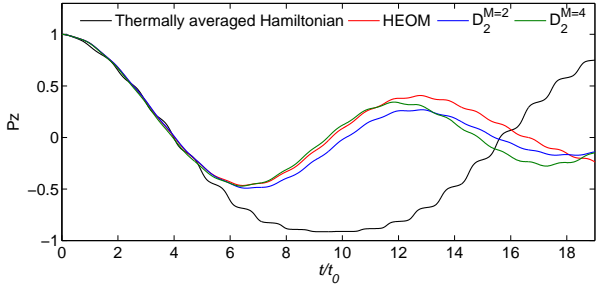}
\caption{$P_{z}(t)$ obtained from averaged Hamiltonian, HEOM method, the $\rm D_2^{M=2}$ {\it Ansatz}, and the $\rm D_2^{M=4}$ {\it Ansatz}. The parameters are $V=-0.05, \lambda=0.5, \beta=0.5$.}
\label{Pzcompare}
\end{figure}

The variational method with importance sampling is simulated by initially employing random number generators to investigate the temperature effects using the multiple Davydov trial wave states. The influence of the temperature on the dynamical behavior is also studied using the method of averaged Hamiltonian with the Davydov D$_1$ {\it Ansatz} developed here (see Eq.~ (\ref{generalized ansatz})) and the HEOM method. Population difference $P_z(t)$ (see Eq.~(\ref{Pz})) obtained from the $\rm D_2^{M=2}$ {\it Ansatz}, the $\rm D_2^{M=4}$ {\it Ansatz}, and the other two methods are plotted in Fig.~\ref{Pzcompare} using parameters $V=-0.05, \lambda=0.5, \beta=0.5$. Unfortunataley, the more complex D$_1$ {\it Ansatz} does not show an improvement over the multi-D$_2$ {\it Ansatz} and the HEOM method, at finite $T$ the distinct damping out of the oscillations is not observed, in contrast to the case of the variational method with importance sampling and the HEOM method. Moreover, with more D$_2$ states used, the results for the variational method with importance sampling come closer to those of the HEOM method.

\subsection{Averaged Hamiltonian}
\label{Thermally averaged Hamiltonian}
According to the papers by Cruzeiro {\it et al.}~\cite{cruz_1988}and by F\"orner \cite{forner_1991}, which are based on earlier work by Davydov and coworkers, temperature effects can be taken into account approximately, by using a generalized Davydov-Ansatz

\begin{eqnarray}
&&\left|\psi_{\nu}\left(t\right)\right\rangle= \nonumber\\
&&A(t)\left|+\right\rangle e^{\left[f(t)\hat{a}^{\dagger}-f^{*}(t)\hat{a}\right]}\left|\nu\right\rangle+B(t)\left|-\right\rangle e^{\left[g(t)\hat{a}^{\dagger}-g^{*}(t)\hat{a}\right]}\left|\nu\right\rangle,\nonumber\\
\end{eqnarray}
\label{generalized ansatz}
with the normalized excited states
\begin{eqnarray}
\left|\nu\right\rangle =\frac{1}{\sqrt{\nu!}}\left(\hat{a}{}^{\dagger}\right)^{\nu}\left|0\right\rangle.
\end{eqnarray}
In this way, one can view the 'thermally averaged state' as a linear combination of all states with a
fixed phonon distribution in the lattice, where the weight factors of the individual states follow Bose-Einstein  statistics. Then we can get a thermally averaged Hamiltonian
\begin{eqnarray}
H_{T}=\sum_{\nu}\rho_{\nu}H_{\nu\nu},
\end{eqnarray}
with
\begin{eqnarray}
\rho_{\nu}=\frac{\left\langle \nu\right|e^{-\beta\omega\hat{a}^{\dagger}\hat{a}}\left|\nu\right\rangle }{\sum_{\nu}\left\langle \nu\right|e^{-\beta\omega\hat{a}^{\dagger}\hat{a}}\left|\nu\right\rangle }=\frac{e^{-\beta\omega\nu}}{Q},
\end{eqnarray}
where $\beta=1/kT$ is proportional to the inverse temperature and
\begin{eqnarray}
H_{\nu\nu}=\left\langle \psi_{\nu}\left(t\right)\right|\hat{H}\left|\psi_{\nu}\left(t\right)\right\rangle,
\end{eqnarray}
in which $\hat{H}$ is given in Eq.~(\ref{SBH}).
Thus,
\begin{eqnarray}
&&H_{T}	=	\frac{\epsilon}{2}\left(\left|A\right|^{2}-\left|B\right|^{2}\right)+V\left(AB^{*}D_{21}+A^{*}BD_{12}\right)\nonumber\\		&&+\omega\left[\left|A\right|^{2}\left(\bar{\nu}+\left|f\right|^{2}\right)+\left|B\right|^{2}\left(\bar{\nu}+\left|g\right|^{2}\right)\right]\nonumber\\
&&-\frac{\lambda}{2}\left[\left|A\right|^{2}\left(f+f^{*}\right)-\left|B\right|^{2}\left(g+g^{*}\right)\right]\nonumber\\
&&+\frac{\phi}{2}\left[AB^{*}\left(g^{*}+f\right)D_{21}+A^{*}B\left(f^{*}+g\right)D_{12}\right],
\end{eqnarray}
where average phonon number is $\bar{\nu}={1}/{(e^{\beta\omega}-1)}$ and
\begin{eqnarray}
&&D_{12}= e^{\left(\bar{\nu}+1\right)f^{*}g+\bar{\nu}g^{*}f-\left(\bar{\nu}+\frac{1}{2}\right)\left(\left|f\right|^{2}+\left|g\right|^{2}\right)},\nonumber\\
&&D_{21}=e^{\left(\bar{\nu}+1\right)g^{*}f+\bar{\nu}f^{*}g-\left(\bar{\nu}+\frac{1}{2}\right)\left(\left|f\right|^{2}+\left|g\right|^{2}\right)}.
\end{eqnarray}
From the Dirac-Frenkel variational principle, we get the equations of motion

\begin{eqnarray}
&&0=i\dot{A}+\frac{i}{2}A\left(f^{*}\dot{f}-\dot{f}^{*}f\right)-\frac{\epsilon}{2}A-\omega A\left(\bar{\nu}+\left|f\right|^{2}\right)\nonumber\\
&&-VBD_{12}-\frac{\lambda}{2}A\left(f+f^{*}\right)-\frac{\phi}{2}B\left(f^{*}+g\right)D_{12},\nonumber\\
&&0=i\dot{B}+\frac{i}{2}B\left(g^{*}\dot{g}-\dot{g}^{*}g\right)+\frac{\epsilon}{2}B-\omega B\left(\bar{\nu}+\left|g\right|^{2}\right)\nonumber\\
&&-VAD_{21}+\frac{\lambda}{2}B\left(g+g^{*}\right)-\frac{\phi}{2}A\left(g^{*}+f\right)D_{21},\nonumber\\
&&0=i\left|A\right|^{2}\dot{f}-\omega\left|A\right|^{2}f-\frac{\lambda}{2}\left|A\right|^{2}\nonumber\\
&&-VAB^{*}\bar{\nu}\left(g-f\right)D_{21}-VA^{*}B\left[\left(\bar{\nu}+1\right)\left(g-f\right)\right]D_{12}\nonumber\\
&&-\frac{\phi}{2}A^{*}BD_{12}-\frac{\phi}{2}AB^{*}\left(g^{*}+f\right)\bar{\nu}\left(g-f\right)D_{21}\nonumber\\
&&-\frac{\phi}{2}A^{*}B\left(f^{*}+g\right)\left(\bar{\nu}+1\right)\left(g-f\right)D_{12},\nonumber\\
&&0=i\left|B\right|^{2}\dot{g}-\omega\left|B\right|^{2}g+\frac{\lambda}{2}\left|B\right|^{2}\nonumber\\
&&-VAB^{*}\left(\bar{\nu}+1\right)\left(f-g\right)D_{21}-VA^{*}B\bar{\nu}\left(f-g\right)D_{12}\nonumber\\
&&-\frac{\phi}{2}AB^{*}D_{21}-\frac{\phi}{2}AB^{*}\left(g^{*}+f\right)\left(\bar{\nu}+1\right)\left(f-g\right)D_{21}\nonumber\\
&&-\frac{\phi}{2}A^{*}B\left(f^{*}+g\right)\bar{\nu}\left(f-g\right)D_{12}.
\end{eqnarray}
In the spin-boson model, physical variables of interest are
\begin{eqnarray}
&&P_{i}\left(t\right)\equiv\left\langle \sigma_{i}\right\rangle =\left\langle D_{s}\right|\sigma_{i}\left|D_{s}\right\rangle , i=x,y,z
\end{eqnarray}
Here $P_z(t)$ describes the population difference. With the above trial state, we obtain
\begin{eqnarray}
&&P_{z}\left(t\right)=\left|A\right|^{2}-\left|B\right|^{2}.
\end{eqnarray}
\label{Pz}
\subsection{Variational method with importance sampling}
\label{Initial displacements based on the Boltzmann distributions}
The variational method with importance sampling is used to obtain the dynamics of the Holstein model with the off-diagonal coupling, where initial phonon displacements are chosen according to the Bose distribution. Using only two sites for simplicity, the Holstein model with the off-diagonal coupling can be reduced to a spin-boson Hamiltonian (Eq.~(\ref{SBH})). We solve the dynamics by variational method using the multi-D$_2$ {\it Ansatz}
\begin{eqnarray}
&&\left|D_2^{M}\left(t\right)\right\rangle= \sum_{i}^{M}A_i(t)\left|+\right\rangle e^{\left[f_i(t)\hat{a}^{\dagger}-f_i^{*}(t)\hat{a}\right]}\left|0\right\rangle\nonumber\\
&&+\sum_{i}^{M}B_i(t)\left|-\right\rangle e^{\left[f_i(t)\hat{a}^{\dagger}-f_i^{*}(t)\hat{a}\right]}\left|0\right\rangle.
\end{eqnarray}
The temperature effects are included by considering the initial displacements based on the Bose distribution \cite{hillery_1984}.
The initial bath can be expressed as
\begin{eqnarray}
\frac{1}{Z_{B}}e{}^{-\beta\omega\hat{a}^{\dagger}\hat{a}}=\int d\alpha^{2}P(\alpha)\left|\alpha\right\rangle \left\langle \alpha\right|,
\end{eqnarray}
where $\left|\alpha\right\rangle \equiv e^{\alpha\hat{a}-\alpha^{*}\hat{a}}\left|0\right\rangle$
and the distribution $P(\alpha)$ is
\begin{eqnarray}
P(\alpha)=\frac{1}{\pi}(e^{\beta\omega}-1)\exp\left(-|\alpha|^{2}(e^{\beta\omega}-1)\right),
\end{eqnarray}
it is shown to be a well behaved Gaussian function and has no singularity. Numerically, let $2\sigma^{2}=1/(e^{\beta\omega}-1)$ and $\alpha=x+iy$,
\begin{eqnarray}
P(\alpha)=\frac{1}{\pi}\frac{1}{2\sigma^{2}}e^{-\frac{x^{2}+y^{2}}{2\sigma^{2}}}=\frac{1}{\sqrt{2\pi}\sigma}e^{-\frac{x^{2}}{2\sigma^{2}}}\frac{1}{\sqrt{2\pi}\sigma}e^{-\frac{y^{2}}{2\sigma^{2}}}.\nonumber\\
\end{eqnarray}\
Then, we can generate the configuration $\alpha$ for the bath according to $P(\alpha)$ by Monte Carlo method. The initial displacements in the trial states is determined by setting $f_{i}\left(t=0\right)=\alpha+\epsilon_{0}$, where a small noise $\epsilon_{0}\in\left[-10^{-2},10^{-2}\right]$ is added to increase the numerical stability. According to the equations of motion obtained from the Dirac-Frenkel variational principle, the dynamics of the system can be obtained. The final result is averaged over enough realizations (more than $50000$) to ensure the convergence of relevant physical quantities. In the same way, initial displacements are also chosen according to the temperature in the fully quantum description of the SSH model.

Following are the corresponded equations of motion
\begin{eqnarray}
&&		-i\sum_{i}\dot{A}{}_{i}S_{ki}\nonumber\\
&&		-\frac{i}{2}\sum_{i}A_{i}\left[-\left(\dot{f}_{i}f_{i}^{\ast}+f_{i}\dot{f}_{i}^{\ast}\right)+2f_{k}^{\ast}\dot{f}_{i}\right]S_{ki}\nonumber\\
&&	=	-\frac{\epsilon}{2}\sum_{i}A_{i}S_{ki}-V\sum_{i}B_{i}S_{ki}-\sum_{i}A_{i}\omega_{0}f_{k}^{\ast}f_{i}S_{ki}\nonumber\\
&&		-\frac{\lambda}{2}\sum_{i}A_{i}\left(f_{i}+f_{k}^{\ast}\right)S_{ki}-\frac{1}{2}\phi\sum_{i}B_{i}\left(f_{i}+f_{k}^{\ast}\right)S_{ki},\nonumber\\
\end{eqnarray}
\begin{eqnarray}
&&		-i\sum_{i}\dot{B}_{i}S_{ki}\nonumber\\
&&		-\frac{i}{2}\sum_{i}B_{i}\left[-\left(\dot{f}_{i}f_{i}^{\ast}+f_{i}\dot{f}_{i}^{\ast}\right)+2f_{k}^{\ast}\dot{f}_{i}\right]S_{ki}\nonumber\\
&&	=	+\frac{\epsilon}{2}\sum_{i}B_{i}S_{ki}-V\sum_{i}A_{i}S_{ki}-\sum_{i}B_{i}\omega_{0}f_{k}^{\ast}f_{i}S_{ki}\nonumber\\
&&		+\frac{\lambda}{2}\sum_{i}B_{i}\left(f_{i}+f_{k}^{\ast}\right)S_{ki}-\frac{1}{2}\phi\sum_{i}A_{i}\left(f_{i}+f_{k}^{\ast}\right)S_{ki},\nonumber\\
\end{eqnarray}
and
\begin{eqnarray}
&&-i\sum_{i}\left(A_{k}^{\ast}\dot{A}_{i}+B_{k}^{\ast}\dot{B}_{i}\right)f_{i}S_{ki}\nonumber\\
&&-i\sum_{i}\left(A_{k}^{\ast}A_{i}+B_{k}^{\ast}B_{i}\right)\dot{f}_{i}S_{ki}\nonumber\\	&&-\frac{i}{2}\sum_{i}\left(A_{k}^{\ast}A_{i}+B_{k}^{\ast}B_{i}\right)f_{i}S_{ki}\left(2f_{k}^{\ast}\dot{f}_{i}-\dot{f}_{i}f_{i}^{\ast}-f_{i}\dot{f}_{i}^{\ast}\right)\nonumber\\
&&=-\frac{\epsilon}{2}\sum_{i}\left(A_{k}^{\ast}A_{i}-B_{k}^{\ast}B_{i}\right)f_{i}S_{ki}\nonumber\\
&&-V\sum_{i}\left(A_{k}^{\ast}B_{i}+B_{k}^{\ast}A_{i}\right)f_{i}S_{ki}\nonumber\\
&&-\sum_{i}\left(A_{k}^{\ast}A_{i}+B_{k}^{\ast}B_{i}\right)\left(\omega_{0}+\omega_{0}f_{k}^{\ast}f_{i}\right)f_{i}S_{ki}\nonumber\\
&&-\frac{\lambda}{2}\sum_{i}\left(A_{k}^{\ast}A_{i}-B_{k}^{\ast}B_{i}\right)S_{ki}\nonumber\\
&&-\frac{\lambda}{2}\sum_{i}\left(A_{k}^{\ast}A_{i}-B_{k}^{\ast}B_{i}\right)f_{i}\left(f_{i}+f_{k}^{\ast}\right)S_{ki}\nonumber\\
&&-\frac{1}{2}\phi\sum_{i}\left(A_{k}^{\ast}B_{i}+B_{k}^{\ast}A_{i}\right)S_{ki}\nonumber\\
&&-\frac{1}{2}\phi\sum_{i}\left(A_{k}^{\ast}B_{i}+B_{k}^{\ast}A_{i}\right)\left(f_{i}+f_{k}^{\ast}\right)f_{i}S_{ki},
\end{eqnarray}

where $S_{ki}=e^{-\frac{1}{2}(f_{k}^{*}f_{k}+f_{i}^{*}f_{i})+f_{k}^{*}f_{i}}$.

---

\end{document}